\documentclass[12pt]{article}
\usepackage{amssymb}
\usepackage{amsmath}
\usepackage{amsfonts}

 \makeatletter
\@addtoreset{equation}{section}

\makeatother
\input{tcilatex}
\begin{document}

\bigskip \thispagestyle{empty}

\begin{center}
\null\vspace{-1cm} \hfill \\[0pt]
\vspace{1cm}{\large \ \ }{\Large \textbf{Liouville Black Hole In A
Noncommutative Space}} \vspace{0.5cm} \bigskip

{\textbf{K. Bilal, A. El Boukili, {M. Nach, } and M.B. Sedra}}\footnote{%
Corresponding author: mysedra@yahoo.fr or msedra@ictp.it}

\vspace{0.5cm}{\emph{Ibn Tofail University, Faculty of Sciences, Physics
Departement, }}

{\emph{(LHESIR), K\'{e}nitra, Morocco.}}

\bigskip

\vspace{1cm} \textbf{Abstract}
\end{center}

The space-noncommutativity adapted to the Liouville black hole theory is
studied in the present work. Among our contributions, we present the
solutions of noncommutative Liouville Black hole equations of motion and
find their classical properties such as the ADM mass, the horizon and the
scalar Ricci curvature.

\bigskip


\section{Liouville black hole}

Consider first the two-dimensional gravity coupled to a Liouville Field $%
\phi $ such that the the action is taken to be \cite{1}%
\begin{equation}
S=S_{G}+S_{L}=S_{G}+\int d^{2}x\sqrt{-g}[b(\nabla \phi )^{2}+\Lambda
e^{-2a\phi }+\gamma \phi R].  \label{1}
\end{equation}

\bigskip The associated field equations are given by 
\begin{equation}
\nabla ^{2}\psi -R=0  \label{2}
\end{equation}%
\begin{equation}
-2b\nabla ^{2}\phi +\gamma R-2a\Lambda e^{-2a\phi }=0  \label{3}
\end{equation}%
and 
\begin{eqnarray}
\frac{1}{2}{\left( \nabla _{\mu }\psi \nabla _{\nu }\psi -\frac{1}{2}g_{\mu
\nu }(\nabla \psi )^{2}\right) } &+&g_{\mu \nu }\nabla ^{2}\psi -\nabla
_{\mu }\nabla _{\nu }\psi \qquad  \notag \\
\quad =8\pi G\left[ -b\left( \nabla _{\mu }\phi \nabla _{\nu }\phi -\frac{%
g_{\mu \nu }}{2}(\nabla \phi )^{2}\right) \right. &-&\left. \gamma \left(
g_{\mu \nu }\nabla ^{2}\phi +\nabla _{\mu }\nabla _{\nu }\phi \right)
\right. \left. +\frac{g_{\mu \nu }}{2}\Lambda e^{-2a\phi }\right]
\end{eqnarray}

\subsection{Exact solutions}

The previous set of equations is shown to reduce to \cite{Mann} 
\begin{equation}
\partial _{+}\partial _{-}(\rho -a\phi )=\frac{8\pi Ga\gamma +4\pi Gb-a^{2}}{%
4\pi G(a\gamma +b)}e^{2(\rho -a\phi )}  \label{5}
\end{equation}%
\textit{i.e.} $\rho -a\phi $ obeys the Liouville equation, provided that $%
a\gamma +b\neq 0$ and $b\neq -4\pi G\gamma ^{2}$.

Using our knowledge of the standard Liouville equation, we can set 
\begin{equation}
\rho -a\phi =\frac{1}{2}\ln \left[ \frac{f_{+}^{\prime }f_{-}^{\prime }}{(%
\frac{A}{K^{2}}-K^{2}f_{+}f_{-})^{2}}\right]  \label{6}
\end{equation}%
where $K$ is a constant of integration and $f_{\pm }$ are arbitrary
functions of the coordinates $x_{\pm }$ and the prime refers to
differentiation by the relevant functional argument. We have also \cite{Mann}
\begin{equation}
\phi =\frac{a-4\pi G\gamma }{4\pi G(a\gamma +b)}\rho +\frac{1}{a}%
(h_{+}+h_{-})+\phi _{0}  \label{7}
\end{equation}%
and%
\begin{equation}
\rho =\xi \ln \left[ \frac{f_{+}^{\prime }f_{-}^{\prime }}{(\frac{A}{K^{2}}%
-K^{2}f_{+}f_{-})^{2}}\right] +h_{+}+h_{-}  \label{8}
\end{equation}%
where $h_{\pm }$ are arbitrary functions of the coordinates $x_{\pm }$
respectively. For convenience an additional constant of integration $\phi
_{0}$ has been retained and we can set%
\begin{equation}
\xi =\frac{2\pi G(a\gamma +b)}{4\pi Gb+8\pi Ga\gamma -a^{2}}  \label{9}
\end{equation}%
yielding the following metric 
\begin{eqnarray}
ds^{2} &=&-e^{2\rho }dx_{+}dx_{-}=-\left[ \frac{f_{+}^{\prime }f_{-}^{\prime
}}{(\frac{A}{K^{2}}-K^{2}f_{+}f_{-})^{2}}\right] ^{2\xi
}e^{2(h_{+}+h_{-})}dx_{+}dx_{-}  \notag \\
&=&-\frac{df_{+}df_{-}}{(\frac{A}{K^{2}}-K^{2}f_{+}f_{-})^{4\xi }}
\label{10}
\end{eqnarray}%
where the last term follows from a suitable choice of $h_{\pm }$. Each of
the metrics (\ref{10}) may be transformed to a static system of coordinates 
\begin{equation}
ds^{2}=-\alpha (x)dt^{2}+\frac{dx^{2}}{\alpha (x)}  \label{11}
\end{equation}%
under an inverse Kruskal-Szekeres transformation. We have to note that there
are three distinct classes of solutions:

$\bigskip $

\textbf{(A)} $\xi \neq 1/4$, $A\neq 0$

In this case, the solution (\ref{11}) is given by 
\begin{eqnarray}
\alpha (x) &=&2Mx-\frac{A}{M^{2}(1-p)^{2}}(2Mx)^{p}=2Mx-Bx_{0}^{2}(\frac{x}{%
x_{0}})^{p}  \notag \\
\phi (x) &=&\frac{2-p}{2a}\ln (2Mx)+\phi _{0}=\frac{2-p}{2a}\ln (x/x_{0})
\label{12} \\
\psi (x) &=&-p\ln (2Mx)+\psi _{0}  \notag
\end{eqnarray}%
where $K=M/(p-1)$ and 
\begin{equation}
p=\frac{4\xi }{4\xi -1}=8\pi G\frac{a\gamma +b}{a^{2}+4\pi Gb}  \label{13}
\end{equation}%
\begin{equation}
B=\frac{4A}{(p-1)^{2}}=\Lambda \frac{(a^{2}+4\pi Gb)^{2}}{(b+4\pi G\gamma
^{2})(4\pi Gb+8\pi Ga\gamma -a^{2})}  \label{14}
\end{equation}%
with $\phi _{0}=\frac{p-2}{2a}\ln (2Mx_{0})$.

\bigskip

\textbf{(B)} $\xi =1/4$, $A\neq 0$

In this case, (\ref{11}) is%
\begin{eqnarray}
\alpha (x) &=&1-Ce^{-2M(x-x_{0})}  \notag \\
\phi (x) &=&\frac{M}{a}(x-x_{0})  \label{15} \\
\psi (x) &=&2Mx+\psi _{0}  \notag
\end{eqnarray}%
where 
\begin{equation}
b=-\frac{a^{2}}{4\pi G},\qquad C\equiv \frac{2\pi G\Lambda a}{M^{2}(a+4\pi
G\gamma )}\text{ \ and }K=M.  \label{16}
\end{equation}%
\bigskip\ \ \ \ 

\textbf{(C)} $\ A=0$

In this case%
\begin{eqnarray}
\alpha (x) &=&2Mx\ln (2Dxx_{0})  \notag \\
\phi (x) &=&\frac{1}{2a}\ln (\frac{4D}{M}x)  \label{17} \\
\psi (x) &=&-2M\ln (\frac{D}{M}x)+\psi _{0}  \notag
\end{eqnarray}%
where $D$ is given by $D=\pi G\Lambda \frac{a}{4\pi G\gamma -a}.$

\subsection{ADM mass, horizon and Ricci scalar curvature}

We summarize here below the essential properties of the Liouville black hole 
\cite{1}

\begin{equation}
\begin{tabular}{|l|c|c|c|}
\hline
& (\textbf{A}) & (\textbf{B}) & \textbf{(C)} \\ \hline
$\alpha (x)$ & $2Mx-Bx_{0}^{2}(\frac{x}{x_{0}})^{p}$ & $1-Ce^{-2M(x-x_{0})}$
& $2Mx\ln (2Dxx_{0})$ \\ \hline
$\mathcal{M}$ & $M\frac{8\pi G\gamma (\pi G{b}+4\pi Ga\gamma )+a^{3}-10\pi
G\gamma a^{2}}{4\pi Ga(a^{2}+4\pi G{b})}$ & $\frac{M}{8\pi G}(1-8\pi G\gamma
/a)$ & $\frac{M}{8\pi G}(1-4\pi G\gamma /a)$ \\ \hline
\textbf{Horizon} $x_{H}$ & $\frac{\mathcal{M}}{\Lambda }\zeta _{1}$ & $\frac{%
a-8\pi G\gamma }{16\pi G\mathcal{M}a}\ln (\frac{\Lambda }{(4\pi G\mathcal{M}%
)^{2}}\zeta _{2})$ & $-\frac{2\mathcal{M}}{\Lambda }e^{2a\phi _{H}}$ \\ 
\hline
\textbf{Curvature} & $Bp(p-1)(\frac{x}{x_{0}})^{p-2}$ & $\frac{8\pi G\Lambda
a}{a+4\pi G\gamma }e^{-2M(x-x_{0})}$ & $-2M/x$ \\ \hline
\end{tabular}
\label{recap}
\end{equation}%
where the, real and positive, ADM-mass $\mathcal{M}$\ is defined as follows%
\begin{equation}
\mathcal{M}=(\frac{\gamma }{4a}-\frac{1}{8\pi G})\alpha ^{\prime }+(\gamma +%
\frac{b}{a})\alpha \phi ^{\prime }-\int \left[ \frac{1}{32\pi G}\left( \frac{%
(\alpha ^{\prime })^{2}-K^{2}}{\alpha }\right) -\frac{b}{2}\alpha (\phi
^{\prime })^{2}-\frac{\gamma }{2}\alpha ^{\prime }\phi ^{\prime }\right]
\label{rich}
\end{equation}

\section{Noncommutativity and Liouville black hole}

The originality of the present work concerns our proposition to study the
effect of the space noncommutativity on the Liouville black hole. For that
we start by introducing the noncommutative variable $\widehat{x}$ subject to
the following noncommutative commutation relations \cite{kr}

\begin{eqnarray}
\left[ \widehat{x}_{\mu },\widehat{x}_{\nu }\right] &=&\widehat{x}_{\mu
}\star \widehat{x}_{\nu }-\widehat{x}_{\nu }\star \widehat{x}_{\mu } \\
&=&i\theta _{\mu \nu }
\end{eqnarray}%
where the non-commutative parameter $\theta _{\mu \nu }$ possesses the
dimension of (length)$^{2}$. The field operators $\widehat{A}(\widehat{x})$
in field theories on NC geometry are the functions of $\widehat{x}_{\mu }$,
where the star-product ($\star $-product) of two fields is given by%
\begin{equation}
A(x)\star B(x)=\left[ \exp \left( \frac{i}{2}\theta _{\mu \nu }\partial
_{\mu }\partial _{\nu }^{\prime }\right) A(x)B(x^{\prime })\right] \mid
_{x=x^{\prime }}.
\end{equation}%
It is implied that we have ordinary commutative relations between
coordinates $\widehat{x}_{\mu }$ and the momentum $\widehat{p}_{\mu }$: 
\begin{equation}
\left[ \widehat{x}_{\mu },\widehat{p}_{\nu }\right] =i\hbar \delta _{\mu \nu
},\quad \left[ \widehat{p}_{\mu },\widehat{p}_{\nu }\right] =0.
\end{equation}%
As $\theta _{\mu \nu }$ is a constant tensor, the Lorentz symmetry is broken
for field theories on NC geometry. It was noted in \cite{kr} that at the
replacement%
\begin{equation}
x_{i}=\widehat{x}_{i}+\frac{1}{2\hbar }\theta _{ij}\widehat{p}_{j},\hspace{%
0.3in}p_{j}=\widehat{p}_{j},
\end{equation}%
we obtain the standard commutation relations 
\begin{equation}
\left[ x_{i},x_{j}\right] =0,\quad \left[ x_{i},p_{j}\right] =i\hbar \delta
_{ij},\quad \left[ p_{i},p_{j}\right] =0.
\end{equation}

\subsection{Noncommutative black hole solutions}

The metric of the Liouville black hole in static system of coordinates is
given by \cite{Mann}%
\begin{equation}
ds^{2}=-\alpha \left( x\right) dt^{2}+\frac{dx^{2}}{\alpha \left( x\right) }
\end{equation}%
We assume here that we can change, in noncommutative space, the coordinate $%
x $ in $\alpha (x)$\ \ by $\hat{x}\bigskip $ as follows:%
\begin{equation}
ds^{2}=-\alpha \left( \hat{x}\right) dt^{2}+\frac{d\hat{x}^{2}}{\alpha
\left( \hat{x}\right) }  \label{(new)}
\end{equation}%
where $\widehat{x}_{i}=x_{i}-\frac{1}{2}\theta _{ij}p_{j}.$ We consider,
after some algebraic computations, three distinct classes of solutions

\begin{itemize}
\item \underline{For $\xi \neq 1/4$, $A\neq 0$}
\end{itemize}

In this case the solution (\ref{(new)}) is 
\begin{eqnarray}
\alpha (\hat{x}) &=&2M\sqrt{\hat{x}\hat{x}}-\frac{A}{M^{2}(1-p)^{2}}(2M\sqrt{%
\hat{x}\hat{x}})^{p}=2M\sqrt{\hat{x}\hat{x}}-Bx_{0}^{2}(\frac{\sqrt{\hat{x}%
\hat{x}}}{x_{0}})^{p}  \notag \\
\phi (\hat{x}) &=&\frac{2-p}{2a}\ln (2M\sqrt{\hat{x}\hat{x}})+\phi _{0}=%
\frac{2-p}{2a}\ln (\sqrt{\hat{x}\hat{x}}/x_{0})  \label{25} \\
\psi (\hat{x}) &=&-p\ln (2M\sqrt{\hat{x}\hat{x}})+\psi _{0}  \notag
\end{eqnarray}%
where $K=M/(p-1)$ and 
\begin{equation}
p=\frac{4\xi }{4\xi -1}=8\pi G\frac{a\gamma +b}{a^{2}+4\pi Gb}  \label{25a}
\end{equation}%
\begin{equation}
B=\frac{4A}{(p-1)^{2}}=\Lambda \frac{(a^{2}+4\pi Gb)^{2}}{(b+4\pi G\gamma
^{2})(4\pi Gb+8\pi Ga\gamma -a^{2})}  \label{26b}
\end{equation}%
with $\phi _{0}=\frac{p-2}{2a}\ln (2Mx_{0})$.

\begin{itemize}
\item \underline{For $\xi =1/4$, $A\neq 0$}
\end{itemize}

In this case (\ref{(new)}) is%
\begin{eqnarray}
\alpha (\hat{x}) &=&1-C\exp \left( -2M\left( \sqrt{\hat{x}\hat{x}}%
-x_{0}\right) \right)  \notag \\
\phi (\hat{x}) &=&\frac{M}{a}(\sqrt{\hat{x}\hat{x}}-x_{0})  \label{29} \\
\psi (\hat{x}) &=&2M\sqrt{\hat{x}\hat{x}}+\psi _{0}  \notag
\end{eqnarray}%
where 
\begin{equation}
b=-\frac{a^{2}}{4\pi G}\qquad C\equiv \frac{2\pi G\Lambda a}{M^{2}(a+4\pi
G\gamma )}  \label{29b}
\end{equation}%
and $K=M$.

\begin{itemize}
\item \underline{For $A=0$}
\end{itemize}

In this case $p=1$ or 
\begin{equation}
b=\frac{a^{2}}{4\pi G}-2a\gamma  \label{29d}
\end{equation}%
and%
\begin{eqnarray}
\alpha (\hat{x}) &=&2M\sqrt{\hat{x}\hat{x}}\ln (2D\sqrt{\hat{x}\hat{x}}x_{0})
\notag \\
\phi (\hat{x}) &=&\frac{1}{2a}\ln (\frac{4D}{M}\sqrt{\hat{x}\hat{x}})
\label{30} \\
\psi (\hat{x}) &=&-2M\ln (\frac{D}{M}\sqrt{\hat{x}\hat{x}})+\psi _{0}  \notag
\end{eqnarray}%
where $\hat{k}^{2}=x_{0}/2$.

\subsection{ADM-Mass, horizon and Ricci \textbf{\textbf{Scalar }}curvature}

This subsection is reserved to a presentation of some properties of
Liouville black holes, namely ADM-Mass, the Horizon and the scalar Ricci
curvature:

\begin{itemize}
\item \underline{\textbf{ADM-Mass}}
\end{itemize}

The ADM-mass in the noncommutative case is defined as follows \cite{Nass}%
\begin{eqnarray}
\mathcal{M} &=&(\frac{\gamma }{4a}-\frac{1}{8\pi G})\alpha ^{\prime
}+(\gamma +\frac{b}{a})\alpha \phi ^{\prime }-\int \left[ \frac{1}{32\pi G}%
\left( \frac{(\alpha ^{\prime })^{2}-K^{2}}{\alpha }\right) \right.  \notag
\\
&&\qquad \qquad \qquad \left. -\frac{b}{2}\alpha (\phi ^{\prime })^{2}-\frac{%
\gamma }{2}\alpha ^{\prime }\phi ^{\prime }\right]
\end{eqnarray}%
except that here $\alpha $ and $\phi $ are defined in terms of
Noncommutative variables $\hat{x}.$ The ADM-Mass will be very useful in the
calculation of thermodynamic properties of Liouville Black Hole such as
Entropy and Temperature.

\underline{For $\xi \neq 1/4$, $A\neq 0$}

In this case the NC ADM-Mass is given by%
\begin{equation}
\mathcal{\hat{M}}=\mathcal{M}-\frac{p\left( 1-p\right) L}{16\hat{K}\pi }%
Bx_{0}^{3-p}x^{p-3}\theta
\end{equation}%
with $\mathcal{M}$\ is given by (\ref{recap},\ref{rich}). For $\theta =0$ we
have $\mathcal{\hat{M}}=\mathcal{M}$

\underline{For $\xi =1/4$, $A\neq 0$}

In this case the NC ADM-Mass is given by%
\begin{equation}
\mathcal{\hat{M}}=\mathcal{M}\exp \left( -2M\Im \left( \theta \right) \right)
\end{equation}%
with%
\begin{equation}
\Im \left( \theta \right) =\frac{x_{0}}{2}\left( \sqrt[\ ]{1+\frac{2L\theta 
}{\left( \frac{1}{2M}\ln c+x_{0}\right) ^{2}}}-1\right)
\end{equation}%
\ and $\mathcal{M}$\ is given by (\ref{recap},\ref{rich}). If $\theta =0$ we
have $\Im \left( 0\right) =0$ $\ $hence$\ \mathcal{\hat{M}}=\mathcal{M}$

\underline{For $A=0$}

In this case the NC ADM-Mass is given by%
\begin{equation}
\mathcal{\hat{M}}=\mathcal{M+}\frac{a-4\pi G\gamma }{8\pi Ga}M\left[ \ln
\left( \lambda (\theta )\right) \right]
\end{equation}%
with%
\begin{equation}
\lambda (\theta )=\frac{1}{2}+\frac{1}{2}\sqrt[\ ]{1+\frac{2L\theta }{\left( 
\frac{4D}{M}\right) ^{2}\exp \left( 4a\phi _{H}\right) }}
\end{equation}%
\ and $\mathcal{M}$\ is given by (\ref{recap},\ref{rich}). If $\theta =0$ we
have $\lambda \left( 0\right) =1$ which implies $\ln \left( \lambda (\theta
)\right) \ $hence$\ \mathcal{\hat{M}}=\mathcal{M}$

\begin{itemize}
\item \underline{\textbf{\textbf{Scalar} Ricci Curvature:}}
\end{itemize}

\underline{\textbf{\ For }$\xi \neq \frac{1}{4},A\neq 0$}: By definition,
the scalar curvature is given by%
\begin{equation}
R=-\frac{d^{2}\alpha \left( \hat{x}\right) }{d\hat{x}^{2}}
\end{equation}%
which gives 
\begin{equation}
R=\frac{-2M}{x-\frac{1}{2}\frac{L\theta }{x}}
\end{equation}%
when $\theta =0$ we obtain the result of the commutative case $R=\frac{2M}{x}%
.$

\underline{\textbf{For }$\xi =\frac{1}{4},A\neq 0$ :} The scalar curvature
in this case is given by:%
\begin{equation}
\alpha \left( \hat{x}\right) =1-c\exp \left( -2M\hat{x}\right)
\end{equation}%
by introducing a double derivation we find%
\begin{equation}
\frac{d^{2}\alpha \left( \hat{x}\right) }{d\hat{x}^{2}}=-4M^{2}c\exp \left(
-2M\hat{x}\right)
\end{equation}%
thus%
\begin{equation}
R=4M^{2}c\exp \left( -2M\left( x-\frac{1}{2}\frac{L\theta }{x}\right) \right)
\end{equation}%
with 
\begin{equation}
c=\frac{2\pi Ga\Lambda }{M^{2}\left( a+4\pi G\gamma \right) }
\end{equation}%
consequently%
\begin{equation}
R=\frac{8\pi Ga\Lambda }{\left( a+4\pi G\gamma \right) }\exp \left(
-2M\left( x-\frac{1}{2}\frac{L\theta }{x}\right) \right)
\end{equation}

\underline{\textbf{For }$\xi \neq \frac{1}{4},A\neq 0$}: The scalar
curvature in this case is given by:%
\begin{equation}
\alpha \left( \hat{x}\right) =2M\hat{x}-Bx_{0}^{2}\left( \frac{\hat{x}}{x_{0}%
}\right) ^{p}
\end{equation}%
consequently%
\begin{equation}
R=p\left( p-1\right) Bx_{0}^{2-p}\left( x-\frac{1}{2}\frac{L\theta }{x}%
\right) ^{p-2}
\end{equation}

\begin{itemize}
\item \underline{\textbf{Horizon of NC\ Liouville Black hole}}
\end{itemize}

\underline{\textbf{For }$\xi =\frac{1}{4},A=0:$}

The horizon of the noncommutative metric (\ref{30}) satisfies the following
condition%
\begin{equation}
2a\phi \left( \hat{x}\right) =\ln \left( \frac{4D}{M}\sqrt{\hat{x}\hat{x}}%
\right)
\end{equation}%
\bigskip By replacing the coordinated $\hat{x}_{i}$ by its value $\hat{x}%
_{i}=x_{i}-\frac{1}{2}\theta _{ij}p_{j}$\ we obtain

\begin{equation}
\sqrt{\left( x_{i}-\frac{1}{2}\theta _{ij}p_{j}\right) \left( x_{i}-\frac{1}{%
2}\theta _{ij}p_{j}\right) }=\frac{M}{4D}\exp \left( 2a\phi _{H}\right) ,
\end{equation}%
which can be reduced simply to

\begin{equation}
x-\frac{1}{2}\frac{\eta \theta }{x}=\frac{M}{4D}\exp \left( 2a\phi
_{H}\right) \ 
\end{equation}%
with the scalar $\eta \theta =x_{i}p_{j}\theta _{ij}$. This gives rise,
after straightforward computations, to the following second order
differential equation\bigskip

\begin{equation}
x^{2}-\left( \frac{M}{4D}\exp \left( 2a\phi \right) \right) x-\frac{1}{2}%
\eta \theta =0\ 
\end{equation}%
\qquad \qquad The allowed solution for the horizon of NC Liouville black
hole is simply given by

\begin{equation}
x_{H}=\frac{M}{4D}\exp \left( 2a\phi _{H}\right) \left( 1+\sqrt[\ ]{1+\frac{%
2\eta \theta }{\left( \frac{M}{4D}\right) ^{2}\exp \left( 4a\phi _{H}\right) 
}}\right) \newline
\end{equation}%
\qquad The relevance of this result is shown once the classical limit is
considered. In fact, the link with the commutative case is provided by
considering $\theta =0$ and thus we find exactly the expresion of the
horizon studied in ref. \cite{Mann}%
\begin{equation}
x_{H}=\frac{M}{2D}\exp \left( 2a\phi _{H}\right)
\end{equation}%
\underline{\textbf{For }$\xi =\frac{1}{4},A\neq 0$ :}

Since the classical metric is given by

\begin{equation*}
\alpha \left( x\right) =1-C\exp \left( -2M\left( x-x_{0}\right) \right)
\end{equation*}%
we can write the metric of the Liouville black hole in a noncommutative
space as%
\begin{eqnarray}
\alpha \left( \hat{x}\right) &=&1-C\exp \left( -2M\left( \hat{x}%
-x_{0}\right) \right)  \notag \\
&=&1-C\exp \left( -2M\sqrt{\hat{x}\hat{x}}\right) \exp \left( 2Mx_{0}\right)
\end{eqnarray}%
The horizon is obtained from the positive solution of the equation $\alpha
\left( \hat{x}\right) =0$ which implies%
\begin{equation}
C\exp \left( -2M\sqrt{\hat{x}\hat{x}}\right) \exp \left( 2Mx_{0}\right) =1
\end{equation}%
So we have the following expression for the horizon%
\begin{equation}
x_{H}=\frac{1}{4M}\ln c+\frac{x_{0}}{2}\left( 1+\sqrt[\ ]{1+\frac{2L\theta }{%
\left( \frac{1}{2M}\ln c+x_{0}\right) ^{2}}}\right)
\end{equation}%
the link with the commutative case is provided by considering $\theta =0$.
In fact, we find exactly the expression of the horizon studied in ref. \cite%
{Mann}$x_{H}=x_{0}+\frac{1}{4M}\ln c$

\bigskip

\textbf{For }$\xi \neq \frac{1}{4},A\neq 0$

In this case we find the horizon of NC Liouville black hole by considering
the field $\phi $ of the metric%
\begin{equation}
\phi =\frac{2-p}{2a}\ln \left( \frac{\sqrt{\hat{x}\hat{x}}}{x_{0}}\right)
\end{equation}%
the development in first order of this equation becomes%
\begin{equation*}
x-\frac{\left( 2-p\right) L\theta }{2x}\left( \frac{1}{x_{0}}\right)
^{2-p}=\exp \left( 2a\phi \right)
\end{equation*}%
which gives the following expression for the horizon%
\begin{equation}
x_{H}=\frac{x_{0}^{2-p}}{2}\exp \left( 2a\phi \right) \left( 1+\sqrt[\ ]{1+%
\frac{2\left( 2-p\right) L\theta }{\left( x_{0}^{2-p}\right) ^{2}\exp \left(
4a\phi \right) }}\right)
\end{equation}%
once again the link with the commutative case is provided by considering $%
\theta =0$, we find exactly the expression of the horizon studied in ref. 
\cite{Mann}%
\begin{equation}
x_{H}=x_{0}^{2-p}\exp \left( 2a\phi \right)
\end{equation}


\begin{thebibliography}{9}
\bibitem{1} R.B. Mann, Phys. Rev. \textbf{D47} (1993) 4438.; R.B. Mann, S.M.
Morsink, A.E. Sikkema and T.G. Steele, Phys. Rev. \textbf{D43} (1991) 3948.

\bibitem{Mann} R.B. Mann, Liouville Black Holes, arXiv:hep-th/9308034v1,
Nucl.Phys. B418 (1994) 231-256

\bibitem{kr} S.I. Kruglov, arXiv:hep-th/0110059v2 24 Nov 2001

\bibitem{Nass} Forough Nasseri, arXiv:hep-th/0508051v1 8 Aug 2005; \ Kourosh
Nozari and Behnaz Fazlpour, arXiv:hep-th/0605109v2 14 Jan 2007; \ Xin-zhou
Li, arXiv:hep-th/0508128v1 18 Aug 2005.

\bibitem{BENS} A. Polyakov. Phys. Lett. B103 (1981) 207.
\end{thebibliography}
\end{document}